\documentclass[12pt]{article}
\usepackage{graphicx, amsmath, amssymb, cite, setspace, color, relsize, bm, bbold, array, hyperref, dsfont, bbold, xcolor, cancel,soul}

\usepackage[top=1 in, bottom=1 in, left=.9 in, right=.9 in]{geometry}

\newcommand{\E}{ |\vec{E}| }

\newcommand{\captionfonts}{\footnotesize} 
\makeatletter
\long\def\@makecaption#1#2{%
  \vskip\abovecaptionskip
  \sbox\@tempboxa{{\captionfonts #1: #2}}%
  \ifdim \wd\@tempboxa >\hsize
    {\captionfonts #1: #2\par}
  \else
    \hbox to\hsize{\hfil\box\@tempboxa\hfil}%
  \fi
  \vskip\belowcaptionskip}
\makeatother

\def\lsim{ \lower .75ex \hbox{$\sim$} \llap{\raise .27ex
\hbox{$<$}} }
\def\gsim{ \lower .75ex \hbox{$\sim$} \llap{\raise .27ex
\hbox{$>$}} }

\def\fnote#1#2{\begingroup\def\thefootnote{#1}\footnote{#2}
     \addtocounter{footnote}{-1}\endgroup}

\let\oldsqrt\sqrt
\def\sqrt{\mathpalette\DHLhksqrt}
\def\DHLhksqrt#1#2{%
\setbox0=\hbox{$#1\oldsqrt{#2\,}$}\dimen0=\ht0
\advance\dimen0-0.2\ht0
\setbox2=\hbox{\vrule height\ht0 depth -\dimen0}%
{\box0\lower0.4pt\box2}}

\begin{document}

\title{Schwinger pair production at nonzero temperatures\\
or in compact directions}

\author{Adam~R.~Brown  \\
 \textit{\small{Physics Department, Stanford University, Stanford, CA 94305, USA}} }
\date{}
\maketitle
\fnote{}{\hspace{-.65cm}email: \tt{mr.adam.brown@gmail.com}}
\vspace{-.95cm}


\begin{abstract}
\noindent Electric fields may decay by quantum tunneling: as  calculated by Schwinger, an electron-positron pair may be summoned from the vacuum. In this paper I calculate the pair-production rate at nonzero temperatures. I find that at high temperatures the decay rate is  dominated by a new instanton that involves both thermal fluctuation and quantum tunneling; this decay is exponentially faster than the  rate in the literature. I also calculate the decay rate when the electric field wraps a compact circle (at zero temperature). The same new instanton also governs this rate: I find that for small circles decay is dominated by a process that drops the electric field by one unit, but does not produce charged particles. 
\end{abstract}

\thispagestyle{empty} 
\newpage

\section{Introduction}
A uniform electric field is classically stable but quantum mechanically unstable. In the semiclassical regime, the dominant decay channel is the nucleation of an electron-positron pair, which discharges a single unit of flux. Heisenberg \& Euler \cite{Heisenberg:1935qt} and then Schwinger  \cite{Schwinger:1951nm} calculated the exponential dependency of the tunneling rate to be
\begin{equation}
\textrm{decay rate}_{T=L^{-1} = 0} \sim \exp \left[ - \frac{1}{\hbar} \frac{\pi m^2}{e \E } \right] . \label{eq:schwingergamma}
\end{equation}
Here $m$ is the positron mass, $e$ is the positron charge, and $\E$ is the electric field strength. 

Barriers that may be traversed by quantum tunneling may also be traversed by thermal fluctuation. I will show that at high temperature  $(T > T_c \equiv \hbar \frac{e \E}{2m})$ the electric field decays by a process in which the electron-positron pair first thermally fluctuates partway up the barrier to nucleation, and only then quantum tunnels through the rest.  For  $T>T_c$ this thermally-assisted quantum tunneling rate,
\begin{equation}
\textrm{decay rate}_{T>T_c} \sim \exp \left[ - \frac{1}{\hbar} \frac{2 m^2}{e |\vec{E}|}  \arcsin \Bigl[ \frac{T_c}{T} \Bigl]  - \frac{m}{T}  \sqrt{ 1 - \frac{T_c^2}{T^2}} \right] , \label{eq:Tgamma}
\end{equation}
is exponentially faster both than the purely quantum Schwinger process (Eq.~\ref{eq:schwingergamma}) and than the purely thermal Boltzmann process (rate $\sim \exp[ - 2m/T]$).

I will also consider the (zero-temperature) decay of an electric field that points down a compact direction of circumference $L$.  I will show that for small circles ($L< L_c   \equiv \frac{2m}{e \E}$)  a real electron-positron pair is not produced; instead, the energy from discharging the flux is dumped into photons. For  $L<L_c$ this is exponentially faster than the Schwinger process:
\begin{equation}
\textrm{decay rate}_{L<L_c}  \sim \exp \left[ - \frac{1}{\hbar}
 \frac{2 m^2}{e |\vec{E}|}  \arcsin \Bigl[ \frac{L }{L_c} \Bigl]  - \frac{m L }{\hbar}  \sqrt{ 1 - \frac{L^2}{L_c^2}} 
\,  \right]  . \label{eq:Lgamma}
\end{equation}

Throughout  I will work in the semiclassical (small $\hbar$) and analogous `semicold' (small $T$) approximations. This is justified if the Compton wavelength $\hbar m^{-1}$ of the electron is short compared to the other length scales in the problem: the electron-positron separation at nucleation $\frac{2m}{e \E}$,  the thermal wavelength $\hbar T^{-1}$, and the circumference of the circle $L$. In the semiclassical regime, decay is slow and exponentially dominated by the tunneling exponent.  
  I will calculate the tunneling exponent at leading order in $\hbar \cdot e \E m^{-2}$ and leading order in $T \cdot m^{-1}$ but at all orders in the ratio of these two expansion parameters $\hbar  e \E m^{-2} /  T  m^{-1} \sim T_c/T$. 

In calculating the rate, there are a number of different formalisms that can be used
 to perform essentially the same mathematical maneuvers \cite{yourcitehere,Coleman:1977py,Brown:1988kg,BlancoPillado:2009di}. 
In the next section I will do a direct WKB calculation; in the appendix I will rederive the same results using an instanton.

\newpage

\begin{figure}[htbp] 
   \centering
   \includegraphics[width=4.2in]{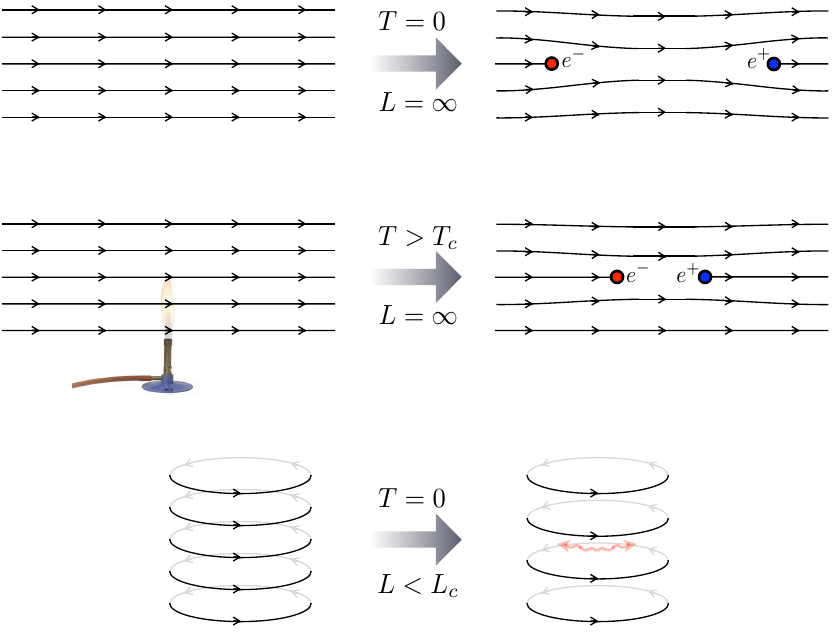} 
   \caption{The three processes considered in this paper. {\bf Top:} standard zero-temperature Schwinger pair production. An electric field line is snipped by the nucleation of an $e^{-} e^{+}$ pair. {\bf Middle:} $T>T_c$  Schwinger pair production has the pair nucleated closer together, using energy extracted from the heat bath. {\bf Bottom:} When the electric field wraps a circle of small circumference $L<L_c$, decay drops the flux  by one unit without producing a real pair; instead the energy is dumped into photons.} 
   \label{fig:allthedecays}
\end{figure}

\section{WKB method} 
A uniform electric field can release its energy by nucleating electron-positron pairs. But there is a barrier. To summon the pair from the vacuum has an upfront cost of $2m$, and this can only be repaid once the pair are far enough apart, $\Delta x \geq 2 \bar{x}_0$ with  
\begin{equation}
 \bar{x}_0  = \frac{m}{e \E} . \label{eq:x0definition}
\end{equation}
%
%
\begin{figure}[h] 
   \centering
   \includegraphics[width=3.5in]{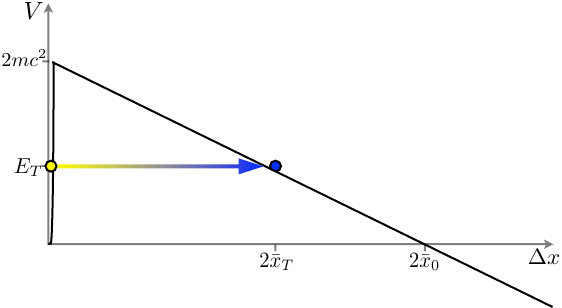} 
   \caption{Traversing the barrier to nucleation. At zero temperature, the pair start colocated and then tunnel to $\Delta x = 2\bar{x}_0$. For $T >T_c$, the pair first thermally fluctuate to an energy $E_T$, then quantum tunnel to $2 \bar{x}_T$.} 
   \label{fig:barrier}
\end{figure}
If the center of nucleation is $x=0$, the positron moves in the potential\footnote{As well as the force from the background $\vec{E}$, the positron also has an O($e^2$) attraction to the electron.
This force is dimension-dependent and small for small $e$; it is easy to include but customary to neglect.} of Fig.~\ref{fig:barrier},  
\begin{gather} 
V_{e^{+}}[x]  = \biggl\{ \begin{array}{ll} 
0 & \textrm{for } x=0 \\ 
m - e \E x  & \textrm{for } x>0,
\end{array} 
\label{VWKBSchwinger}
\end{gather}
and the electron moves towards negative $x$ is a similar potential. To make the positron costs $m$, but the farther it goes in $x$, the more of that energy it recovers from the electric field. 

At zero temperature, the barrier is traversed by tunneling. 
The WKB tunneling suppression $\exp [- I_{\textrm{e}^{+}}/\hbar]$ can be calculated by integrating the positron's (imaginary) momentum across the barrier; for a positron with no initial energy the relativistic dispersion relation then gives \cite{Brown:2007vha} 
\begin{equation}
I_{\textrm{e}^{+}} = 2i  \int_0^{\bar{x}_0} dx \, p = 2i  \int_0^{\bar{x}_0} dx \, \sqrt{(V-m)^2 - m^2} = 2 \int_0^{\bar{x}_0} dx \sqrt{ V(x) [2m - V(x)] }. \label{eq:relativisticWKB}
\end{equation}
(This reduces to the standard $2\int \sqrt{2mV}$ in the non-relativistic limit $V \ll m$; note  that for $V > m$ higher barriers mean faster tunneling \cite{Brown:2007vha,KleinParadox}.) Applying Eq.~\ref{eq:relativisticWKB} to Eq.~\ref{VWKBSchwinger}  and including both $I_{e^+}$ and $I_{e^-}$ recovers the Schwinger rate Eq.~\ref{eq:schwingergamma}. 

\subsection{$\bf{T > T_c}$}
Now let's turn on a temperature. Decay will proceed by a three step process: first absorb energy from the heat bath, then quantum tunnel at fixed energy, then classically roll down the potential \cite{Affleck:1980ac}. To conserve energy, a positron endowed by the heat bath with an energy $E_i$ need only tunnel out to 
\begin{equation}
\bar{x}_{E_i} =  \frac{m - E_i}{e \E} \ \leq  \ \bar{x}_0  .\label{eq:x0ofE0}
\end{equation}
The optimal value of $E_i$ is determined by the interplay of  a thermal (Boltzmann) suppression that wants $E_i$ to be small, and a quantum (WKB) suppression that wants $E_i$ to be big:
\begin{eqnarray}
\textrm{rate}_{e^+}(E_i,T) & = & \exp\biggl[ - \frac{E_i}{T} - \frac{2}{\hbar} \int_{0}^{{\bar{x}_{E_i}}} dx \sqrt{[V-  E_i][2m - (V-E_i)]} \, \biggl]   \\
& = & \exp \biggl[  - \frac{E_i}{T} - \frac{2}{\hbar} \left(  \frac{m^2}{2 e \E} \arccos \left[ \frac{E_i}{m} \right]     -\frac{{E_i} \sqrt{ m^2-{E_i}^2}}{2  e \E } \right) \biggl] . \label{eq:rateE0notyetT}  
\end{eqnarray}
For $T<T_c \equiv \hbar \frac{e \E}{2m}$ the decay rate is maximized at $E_i = 0$ and we recover the Schwinger result. For $T>T_c$ the optimal tradeoff between the quantum and thermal factors is given by $\partial_{E_i} \textrm{rate}_{e^+}(E_i,T) = 0$ as 
\begin{equation}
2 E_i \Bigl|_\textrm{fastest}  = E_T \equiv  2m \sqrt{ 1 - \frac{T_c^2}{T^2} } . \label{eq:E0}
\end{equation}
This energy has been extracted from the heat bath. Applying Eq.~\ref{eq:E0} to 
Eq.~\ref{eq:rateE0notyetT} successfully recovers the decay rate of Eq.~\ref{eq:Tgamma}.

(Indeed, this analysis also allows us to apportion the total decay suppression between the Boltzmann factor and the WKB factor as $I_{\textrm{total}} =  I_\textrm{thermal}+ I_\textrm{quantum}$ where 
$I_{\textrm{thermal}}  =  \frac{2E_i}{T}  = \frac{ 2m}{T} \sqrt{1 - \frac{T_c^2}{T^2}}$ and $ I_\textrm{quantum}    =   \frac{2 m^2}{e |\vec{E}|}  \arcsin \Bigl[ \frac{T_c}{T} \Bigl]  - \frac{m}{T}  \sqrt{ 1 - \frac{T_c^2}{T^2}}$.)

At high temperature 
the decay rate is 
\begin{equation}
\textrm{rate}_{T \gg T_c} = \exp \left[ - \frac{2 m}{T} \left( 1 -  \frac{1}{6} \frac{T_c^2}{T^2} + \ldots \right)  \right] .
\end{equation}
The leading term in the exponent is the Boltzmann suppression of making 
a pair.

Schwinger pair production at nonzero temperature has been considered before \cite{Loewe:1991mn,Elmfors:1994fw,Hallin:1994ad,Ganguly:1995mi,Gies:1998vt,Gies:1999vb,Gies:2000ae,Kim:2007ra,Kim:2008em,Medina:2015qzc} without discovering this process\footnote{For example, according to the result advanced in \cite{Medina:2015qzc}, as the temperature goes up the pair-production rate goes down; my rate Eq.~\ref{eq:Tgamma} has the opposite behaviour. Furthermore, since the tunneling rate is only modified for $T>T_c$, the correction is invisible to the analysis of  \cite{Elmfors:1994fw,Gies:1998vt}.} or the rate of Eq.~\ref{eq:Tgamma}.

\subsection{$\bf{L<L_c}$} \label{sec:LlessthanLc}

Now consider an electric field at zero temperature that points down a compact  direction of circumference $L$. The electric field lines form closed loops, as in the bottom panel of Fig.~\ref{fig:allthedecays}. This configuration is a perturbatively stable solution to the sourceless Maxwell equations, but it will decay nonpeturbatively via quantum tunneling. 

For $L > L_c \equiv 2 \bar{x}_0$ the electric field decays by standard Schwinger pair production. In this parameter regime the nucleation of the electron-positron pair is unaffected by the compactness of the electric-field direction, so an electron and a positron are produced at rest at a separation $\Delta x = 2 \bar{x}_0$. The electric field then forces the pair to classically accelerate apart. Since the direction in which they are accelerating is compact, the electron and positron will meet again, and then either collide with each other and annihilate into photons, or miss each other and go round again, discharging a unit of flux on each lap \cite{Kleban:2011cs}.

For $L<L_c \equiv 2 \bar{x}_0$ the energy available from unwrapping a single unit of flux from the compact direction
\begin{equation}
\textrm{Energy Released} = e \E L < 2m
\end{equation} 
 is insufficient to repay the $2m$ required to fabricate a pair. So the Schwinger formula Eq.~\ref{eq:x0definition} predicts that in order to conserve energy the electron-positron pair, having tunneled  to $\Delta x = L$, must continue tunneling all the way to $\Delta x = 2 \bar{x}_0$. But there is another way to conserve energy: finding themselves reunited at $\Delta x = L$, the electron and positron can immediately annihilate. Because this process does not create an on-shell electron-positron pair, there is no $2m$ to pay and energy can be conserved without tunneling all the way to $\Delta x = 2 \bar{x}_0$. 
Since this process doesn't require any further tunneling beyond $\Delta x = L$, this process is exponentially faster than standard Schwinger pair production; tunneling from $\Delta x = 0$ to $\Delta x = L$ in Eq.~\ref{eq:relativisticWKB} recovers the decay rate of Eq.~\ref{eq:Lgamma}.
Thus for $L<L_c \equiv 2 \bar{x}_0$ the dominant decay process is a virtual pair circumnavigating the circle once and then annihilating into photons; the flux quantum jumps down by one unit without ever creating an on-shell electron-positron pair.

For  small circles 
the decay rate, Eq.~\ref{eq:Lgamma}, becomes 
\begin{equation}
\textrm{rate}_{L \ll L_c} = \exp \left[ - \frac{2 m L }{\hbar} \left( 1 -  \frac{1}{6} \frac{L^2}{L_c^2} + \ldots \right)  \right] .
\end{equation}

\section{Discussion}

In this paper, we have explored two complications we can add to the standard Schwinger story about tunneling in a uniform electric field: we have added a temperature, and we have compactified the direction down which the electric field points. In both cases, the rates are exponentially faster than that derived by Schwinger. 

Associated with the faster processes is some new phenomenology. In the thermal case, we found that the total energy of the electric field plus particles is not conserved. Energy is taken from the heat bath during the nucleation process that is never returned. In the compact case, the deviation is  more dramatic. Unlike in the Schwinger process, no real pair is created. Instead a virtual pair mediates the unwrapping of a single unit of flux, 
but the pair annihilates before the tunneling process is complete, dumping the liberated energy into photons. In this process, not only is the generated current quantized in space, it is also quantized in time.

Let's compare thermally-assisted quantum tunneling in the potential of Fig.~\ref{fig:barrier} to thermally-assisted quantum tunneling in two better-studied examples \cite{Affleck:1980ac,Linde:1980tt,Coleman:1980aw,Hackworth:2004xb,Batra:2006rz,Brown:2007sd,Brown:2014rka}; we'll find  differences caused by the non-analyticity of the potential at $\Delta x = 0$. The first well-studied example is either quantum mechanics or quantum field theory when the barrier to be traversed is smooth. In that case,  even at arbitrarily low nonzero temperatures the dominant process does not tunnel the whole way but instead receives at least a small thermal assist from the heat bath ($E_T > 0$); and furthermore above a critical temperature the process is purely thermal, meaning it fluctuates straight to the top of the barrier and doesn't quantum tunnel at all
\cite{Hawking:1981fz,Jensen:1983ac}. The second well-studied example is tunneling of three-dimensional quantum fields whose barriers have been rendered non-analytic by taking the thin-wall limit. In this case there is a first-order transition in the decay rate: at a critical temperature, dominance jumps from purely quantum (spherical instantons) to purely thermal (cylindrical instantons). Thermal Schwinger pair production is different from both of these examples. Instead, at low temperature ($T<T_c$) the dominant process is purely quantum (as in the thin-wall three-dimensional case), but there is then a higher-order transition in the decay rate, with the decay acquiring an at-first-small thermal assist; furthermore, no matter how high the temperature the process never becomes purely thermal.

The non-analyticity of the tunneling exponent at $T = T_c$ is an artifact of having taken the semiclassical limit, and O$(\hbar m^{-1})$ corrections will smooth it to a finitely sharp crossover. (See for example the discussion in Sec.~4 of \cite{Brown:2011um}.)

One might have worried that thermally-assisted quantum tunneling is the answer to an ill-posed question, since even with $\hbar = 0$ the electric field is discharged by the Boltzmann disassociation of electrons and positrons. But Eq.~\ref{eq:Tgamma} proves this concern is unfounded. Thermally-assisted quantum pair production can be cleanly distinguished from purely thermal pair production because the rate for the first (Eq.~\ref{eq:Tgamma}) is exponentially faster than the rate for the second ($\exp[ - 2m/T]$).  
(This is essentially the same reason as  assures the well-posedness of \cite{Affleck:1980ac,Linde:1980tt,Coleman:1980aw}.)


Equations~\ref{eq:Tgamma} and \ref{eq:Lgamma} give the exponential contribution to the decay rate; in the semiclassical regime this is exponentially the most important contribution. A next step would be to calculate the leading contribution to the  prefactor. This can be done using the instanton discussed in the appendix by calculating the determinant of the matrix of small perturbations \cite{Callan:1977pt}. 

This determinant is negative because, as befits a tunneling instanton, one of the eigenvalues is negative. Eigenmodes that leave the vertex unchanged are all positive \cite{Medina:2015qzc}, but the eigenmode that ``reconnects'' the vertex is negative. This mode deforms the sharp corner at which the two walls meet, smoothing it into an avoided crossing with lower Euclidean action. 


If both $T > T_c$ {and} $L<L_c$ then both Euclidean directions are compact. For $\beta < L$ the dominant process is  thermal nucleation, as in Sec.~2. For $\beta > L$ the dominant process is a quantum jump in the flux without producing charged particles, as in Sec.~3. There is also a process in which the quantum jump receives a thermal assist (governed by a Euclidean solution with four conjoined less-than-quarter-circular arcs) but it can be shown that this is subdominant. Giving the heat bath momentum in the compact direction would induce a chemical potential and make the  Euclidean space in which the instanton lives deform from a rectangle to a parallelogram.

The two rates Eqs.~\ref{eq:Tgamma} and \ref{eq:Lgamma} are related by the duality $\beta \leftrightarrow L$. Mathematically, this is because the same  instanton controls both decays, only with relabelled axes. Since this relates large and small values of $L/\beta$, this is a high-temperature/low-temperature duality.

Experimentalists have sought to detect Schwinger pair production and measure the decay rate, Eq.~\ref{eq:schwingergamma}, using graphene \cite{graphene,Allor:2007ei}. Both Eq.~\ref{eq:Tgamma} and Eq.~\ref{eq:Lgamma} are of the right order to be detectable with state-of-the-art techniques. For example, consider bilayer graphene with a mass-gap of order 200$meV$, a sample size of 10$\mu$, a total voltage drop of a few mass-gaps, and at a temperature 300$K$: decay is semiclassical and prompt enough to be detectable, and  since $T \ \gsim \ T_c$ this is a good system in which to test Eq.~\ref{eq:Tgamma}. Perhaps Eq.~\ref{eq:Lgamma} could be  tested with carbon nanotubes.

\section*{Acknowledgements}
Thank you to Matt Reece for early collaboration on this project, and thank you to Erick Weinberg, Don Page, and David Goldhaber-Gordon. Thank you to the Aspen Center for Physics and the NSF Grant \#1066293 for hospitality during part of the production of this paper.

\appendix

\section{Instanton Method} \label{sec:instantonmethod}

In  Sec.~2, we calculated the decay rate using the WKB formalism. In this section, I will provide an alternative perspective by rederiving the same result using an instanton.

\subsection{$\mathbf{T = L^{-1} = 0}$}

Schwinger pair production can be thought of as the decay of the false vacuum of a 1+1-dimensional quantum field. In this language, pair production is the nucleation of a bubble of true vacuum \cite{Coleman:1977py,Brown:1988kg,BlancoPillado:2009di}. The interior of the bubble has reduced electric field, and hence a lower energy density, $\Delta (\frac{1}{2} \E^2) =- e \E + \textrm{O}(e^2)$; the surface of the bubble is the charged particle, which forms a thin wall of thickness $\hbar m^{-1}$ and tension $m$. Thus for Schwinger pair production, the semiclassical and thin-wall regimes coincide. 

Since the electric field is uniform, the perpendicular spatial directions are passive spectators that can be integrated out. Consequently, we can use the results of 1+1-dimensional field theory, including Eqs.~\ref{eq:schwingergamma}-\ref{eq:Lgamma}, in any number of spatial dimensions\footnote{Ironically, the only number of dimensions in which the results of this paper cannot be trusted is 1+1, since QED in 1+1 dimensions has no dynamical photons, and so no dense spacing of energy levels to underwrite the semiclassical approximation. The spectator degrees of freedom do not directly contribute to the WKB tunneling rate---rather they are required for the WKB approximation to be valid at all. (For an analogous consideration in  conventional one-dimensional quantum mechanical tunneling, see \cite{Nieto:1985ws}).}.

The Euclidean action for a bubble of true vacuum with surface at $x(\tau)$ is
\begin{equation}
I_\textrm{Euclidean} = m \times \textrm{perimeter} - e \E \times \textrm{area} = \int  d \tau \left( m \sqrt{ 1+ \dot{x}^2} - e \E x  \right). \label{eq:euclideanaction}
\end{equation}
(Not coincidentally, this is also the  action  of a relativistic particle in a uniform electric field.) 
The tunneling  instanton is a saddle point of the Euclidean action with a single negative mode. The Euler-Lagrange equation tells us that, for these boundary conditions, the instanton is a circular bubble of radius $\bar{x}_0$ (which w.l.o.g.~we may center at the origin) 
\begin{equation}
x^2 + \tau^2 = \bar{x}_0^2. 
\end{equation}
The exponential of the instanton action $\exp [ -  I_\textrm{Euclidean} /{\hbar} ] = \exp [ -  ( m \times   2 \pi \bar{x}_0 -  e \E \times \pi \bar{x}_0^2 ) / {\hbar}]$ gives the decay rate, Eq.~\ref{eq:schwingergamma}. 
The classical Lorentzian evolution after nucleation is given by analytically continuing 
$\tau \rightarrow i t$:
\begin{equation}
x^2 - t^2 =  \bar{x}_0^2 .
\end{equation}
The electron and positron are nucleated at rest at $x = \mp {\bar{x}_0}$, and then accelerate apart.

\begin{figure}[htbp] 
   \centering
   \includegraphics[width=2.5in]{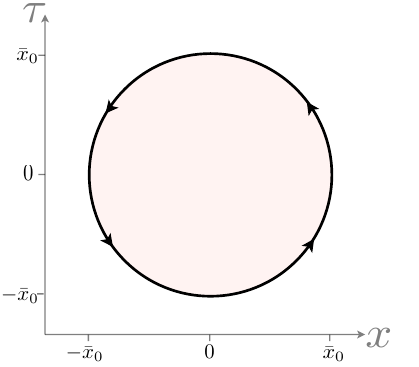} 
   \caption{The Schwinger pair-production  instanton is a bubble of radius $\bar{x}_0 = {m}/{e \E}$. Inside the bubble (shaded) the electric field is reduced and so the energy density is lower. 
  The slice through $\tau = 0$ gives the $t=0$ state immediately following nucleation: a momentarily stationary electron and positron separated by $2\bar{x}_0$.}
\end{figure}

\subsection{$\mathbf{T > T_c}$}  \label{sec:withinstantonTbiggerthanTc}
At nonzero temperature, a particle may thermally fluctuate partway up the barrier before tunneling through the rest. The nonzero temperature makes Euclidean time compact  \cite{Affleck:1980ac,Linde:1980tt} with period $\beta \equiv \frac{\hbar}{T}$,
\begin{equation}
I_\textrm{Euclidean} = \int_{- \frac{\beta}{2}}^{\frac{\beta}{2}}   d \tau \left( m \sqrt{ 1+ \dot{x}^2} - e \E x \right). \label{eq:euclideanaction}
\end{equation}

For $T<T_c$ (equivalently $\beta > \beta_c \equiv 2 \bar{x}_0$) the complete Schwinger bubble still fits into the compact direction and the temperature has no effect on the tunneling exponent. 

\begin{figure}[htbp] 
   \centering
   \includegraphics[width=5.5in]{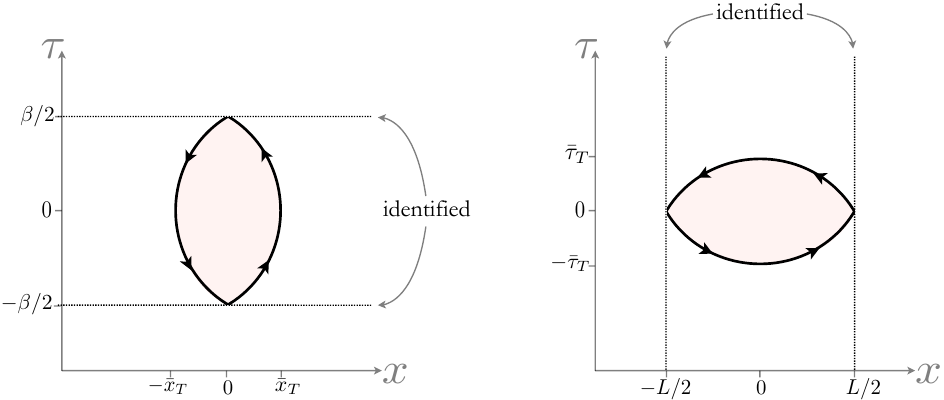} 
   \caption{{\bf Left}: for the nonzero-temperature instanton, Euclidean time is compact. Two less-than-semicircular arcs, each of radius of curvature $\bar{x}_0$, meet at $x = 0, \tau = \pm \beta/2$. Inside the bubble (shaded) the electric field is reduced  and so the energy density is lower.   The slice through $\tau = 0$ gives the $t=0$ state immediately following nucleation: a momentarily stationary electron and positron separated by $2\bar{x}_T$. {\bf Right}: the same instanton does double duty, also describing the decay of a zero-temperature electric field pointing down a compact spatial direction of circumference $L$. Instead of producing an electron-positron pair, tunneling instead dumps the discharged energy into photons.}
   \label{fig:finiteinstantons}
\end{figure}

For $T>T_c$ the circular bubble no longer fits. Instead, the dominant instanton is a lens-shaped bubble\footnote{This same lens-shaped extremal surface has appeared in the literature before, though without the connection being made to Schwinger pair production at finite temperature or with compact directions \cite{Selivanov:1985vt,Selivanov:1986tu,Ivlev:1987zz,Schutzhold:2008pz,Schneider:2014mla}.} bounded by two less-than-semicircular arcs, as shown in Fig.~\ref{fig:finiteinstantons}. Since the equations of motion are locally the same as for $T=0$, the bubble walls must still be segments of a circle of radius $\bar{x}_0$. The bubble has maximum width at $\tau = 0$ where $x = \pm \bar{x}_T$ with 
\begin{equation}
\bar{x}_T = \bar{x}_0 \left( 1 - \sqrt{ 1 - \frac{T_c^2}{T^2} } \right)  , \label{eq:xT}
\end{equation}
and narrows to a vertex\footnote{That the vertex itself satisfies the equation of motion follows from the two $Z_2$ reflection symmetries, which guarantee that the forces on the vertex cancel. Within a Compton wavelength of the vertex the electron-positron   bilateral interaction is important---this produces a force that causes the trajectories to `bounce', but does not contribute to the action beyond O($e^2$) since the action has no extrinsic curvature terms.}  at $x = 0, \tau = \pm \beta/2$.  The exponential of the instanton action $\exp [ -  I_\textrm{Euclidean} /{\hbar}] = \exp[-(m \times \textrm{perimeter} - e \E \times \textrm{area})/\hbar]$ gives the decay rate,  Eq.~\ref{eq:Tgamma}.

The classical Lorentzian evolution after nucleation is given by analytically continuing $\tau \rightarrow i t$. The electron and positron are initially at rest at $x= \mp \bar{x}_T$, and then accelerate apart with $\alpha = \bar{x}_0^{-1}$. Since $\bar{x}_T < \bar{x}_0$, the trajectory of the nucleated pair breaks the boost symmetry and gives a preferred frame. This preferred frame is inherited from the rest frame of the heat bath. 

A stationary electron-positron pair separated by $2 \bar{x}_T$ has energy given by $E_T$ from Eq.~\ref{eq:E0}. The instanton thus automatically calculates the optimal value of $E_i$ that gives the fastest decay path \cite{Affleck:1980ac}.

The purely thermal solution with two conjoined thin walls has extra negative modes, analogous to \cite{instability}, and never dominates tunneling.

\subsection{$\mathbf{L < L_c}$}
Let's re-derive the result of Sec.~\ref{sec:LlessthanLc} using an instanton. Pleasingly, we will find that the pertinent instanton is the same one that also mediated the thermal decay of Sec.~\ref{sec:withinstantonTbiggerthanTc}. 

Consider the zero-temperature decay of a constant electric field that points down a compact direction of circumference $L$. The instanton that describes  this process is shown in Fig.~\ref{fig:finiteinstantons}; the lens-shaped bubble extends as far as $\tau = \pm \bar{\tau}_T \equiv  \pm \bar{x}_0 (1 - \sqrt{1 - L^2/L_c^2})$. 

To interpret this instanton, recall that the classical configuration immediately after tunneling ($t=0$) is given by the $\tau = 0$ time slice through the instanton. As shown in Fig.~\ref{fig:finiteinstantons}, this process drops a unit of flux throughout the entire space. The $t=0$ time slice also includes the vertex, confirming the interpretation of Sec.~\ref{sec:LlessthanLc} which is that after tunneling there are no electrons or positrons left over, and instead the flux has quantum jumped down by one unit without the production of on-shell charged particles. 

 The finite-$L$ instanton is the same mathematical solution as the finite-$\beta$ instanton, only now with the two compact Euclidean directions swapped. Since the value of the Euclidean action is insensitive to how we label the axes, the rate Eq.~\ref{eq:Lgamma} follows directly by substituting $\beta \rightarrow L$ into the thermal rate Eq.~\ref{eq:Tgamma}. There is thus a high-temperature/low-temperature duality between the two decay processes considered in this paper. 
%
%
%

(A similar instanton may be used to calculate the discharge rate of an electric field between two capacitive plates, a problem considered without the use of instantons in \cite{Wang:1988ct}.)\\



\begin{thebibliography}{99}

\bibitem{Heisenberg:1935qt} 
  W.~Heisenberg and H.~Euler,
  ``Consequences of Dirac's theory of positrons,''
  Z.\ Phys.\  {\bf 98}, 714 (1936)
  [physics/0605038].
  %
  %
  %
  %
\bibitem{Schwinger:1951nm} 
  J.~S.~Schwinger,
  ``On gauge invariance and vacuum polarization,''
  Phys.\ Rev.\  {\bf 82}, 664 (1951).
  
\bibitem{yourcitehere}
  I.~K.~Affleck, O.~Alvarez and N.~S.~Manton,
  ``Pair Production at Strong Coupling in Weak External Fields,''
  Nucl.\ Phys.\ B {\bf 197}, 509 (1982);
%
%
%
  I.~K.~Affleck and N.~S.~Manton,
  ``Monopole Pair Production in a Magnetic Field,''
  Nucl.\ Phys.\ B {\bf 194}, 38 (1982);
%
%
  S.~P.~Kim and D.~N.~Page,
  ``Schwinger pair production via instantons in a strong electric field,''
  Phys.\ Rev.\ D {\bf 65}, 105002 (2002)
  [hep-th/0005078];
  %
  %
  %
  S.~P.~Kim and D.~N.~Page,
  ``Schwinger pair production in electric and magnetic fields,''
  Phys.\ Rev.\ D {\bf 73}, 065020 (2006)
  [hep-th/0301132];
%
%
%
   %
  G.~V.~Dunne,
``Heisenberg-Euler effective Lagrangians: Basics and extensions,''
  In *Shifman, M. (ed.) et al.: From fields to strings, vol. 1* 445-522
  [hep-th/0406216];
%
%
%
  G.~V.~Dunne and C.~Schubert,
  ``Worldline instantons and pair production in inhomogeneous fields,''
  Phys.\ Rev.\ D {\bf 72}, 105004 (2005)
  [hep-th/0507174];
  %
  %
%
%
  C.~Schubert,
  ``QED in the worldline representation,''
  AIP Conf.\ Proc.\  {\bf 917}, 178 (2007)
  [hep-th/0703186];
  %
    %
  T.~D.~Cohen and D.~A.~McGady,
  ``The Schwinger mechanism revisited,''
  Phys.\ Rev.\ D {\bf 78}, 036008 (2008)
  [arXiv:0807.1117 [hep-ph]];
  %
  S.~P.~Kim and D.~N.~Page,
  ``Schwinger Pair Production in dS(2) and AdS(2),''
  Phys.\ Rev.\ D {\bf 78}, 103517 (2008)
  [arXiv:0803.2555 [hep-th]].
  


  

\bibitem{Coleman:1977py} 
  S.~R.~Coleman,
  ``The Fate of the False Vacuum. 1. Semiclassical Theory,''
  Phys.\ Rev.\ D {\bf 15}, 2929 (1977)
  [Erratum-ibid.\ D {\bf 16}, 1248 (1977)].


\bibitem{Brown:1988kg} 
  J.~D.~Brown and C.~Teitelboim,
  ``Neutralization of the Cosmological Constant by Membrane Creation,''
  Nucl.\ Phys.\ B {\bf 297}, 787 (1988).

    
\bibitem{BlancoPillado:2009di} 
  J.~J.~Blanco-Pillado, D.~Schwartz-Perlov and A.~Vilenkin,
``Quantum Tunneling in Flux Compactifications,''
  JCAP {\bf 0912}, 006 (2009)
  [arXiv:0904.3106 [hep-th]].



\bibitem{Brown:2007vha} 
  A.~R.~Brown,
  ``Brane tunneling and virtual brane-antibrane pairs,''
  PoS CARGESE {\bf 2007}, 022 (2007)
  [arXiv:0709.3532 [hep-th]].
  

\bibitem{KleinParadox}
O.~Klein, ``Die reflexion von elektronen an einem potentialsprung nach der relativistischen dynamik von Dirac.''
 Z. Phys. {\bf 53}, 157Ð165 (1929); 
W.~Greiner, B.~Mueller, and J.~Rafelski, ``Quantum Electrodynamics of Strong Fields'', (Springer, Berlin, 1985); 
A.~A.~Grib, S.~G.~Mamayev, and  V.~M.~Mostepanenko,  ``Vacuum Effects in Strong Fields'' (Friedmann, St-Petersburg, 1994); 
R. K. Su, G.~C.~Siu, and X.~Chou, ``Barrier penetration and Klein paradox'', J. Phys. A {\bf 26}, 1001Ð1005 (1993); 
N.~Dombey and A.~Calogeracos, ``Seventy years of the Klein paradox'', Phys. Rep. {\bf 315}, 41Ð58 (1999); A.~Calogeracos and N.~Dombey, ``History and physics of the Klein paradox'', Contemp. Phys. {\bf 40}, 313Ð321 (1999); 
P.~Krekora, Q.~Su, and R.~Grobe, ``Klein paradox in spatial and temporal resolution'' Phys. Rev. Lett. {\bf 92}, 040406 (2004); 
  A.~R.~Brown, S.~Sarangi, B.~Shlaer and A.~Weltman,
  ``A Wrinkle in Coleman-De Luccia,''
  Phys.\ Rev.\ Lett.\  {\bf 99}, 161601 (2007)
  [arXiv:0706.0485 [hep-th]].

\bibitem{Affleck:1980ac} 
  I.~Affleck,
  ``Quantum Statistical Metastability,''
  Phys.\ Rev.\ Lett.\  {\bf 46}, 388 (1981).
%


  


\bibitem{Brown:2011um} 
  A.~R.~Brown and A.~Dahlen,
  ``The Case of the Disappearing Instanton,''
  Phys.\ Rev.\ D {\bf 84}, 105004 (2011)
  [arXiv:1106.0527 [hep-th]].

\bibitem{Loewe:1991mn} 
  M.~Loewe and J.~C.~Rojas,
  ``Thermal effects and the effective action of quantum electrodynamics,''
  Phys.\ Rev.\ D {\bf 46}, 2689 (1992).

\bibitem{Elmfors:1994fw} 
  P.~Elmfors and B.~S.~Skagerstam,
  ``Electromagnetic fields in a thermal background,''
  Phys.\ Lett.\ B {\bf 348}, 141 (1995)
  [Phys.\ Lett.\ B {\bf 376}, 330 (1996)]
  [hep-th/9404106].
  %
  
\bibitem{Hallin:1994ad} 
  J.~Hallin and P.~Liljenberg,
  ``Fermionic and bosonic pair creation in an external electric field at finite temperature using the functional Schrodinger representation,''
  Phys.\ Rev.\ D {\bf 52}, 1150 (1995)
  [hep-th/9412188].
  
\bibitem{Ganguly:1995mi} 
  A.~K.~Ganguly, J.~C.~Parikh and P.~K.~Kaw,
  ``Thermal tunneling of q anti-q pairs in A-A collisions,''
  Phys.\ Rev.\ C {\bf 51}, 2091 (1995).

\bibitem{Gies:1998vt} 
  H.~Gies,
  ``QED effective action at finite temperature,''
  Phys.\ Rev.\ D {\bf 60}, 105002 (1999)
  [hep-ph/9812436].
  
\bibitem{Gies:1999vb} 
  H.~Gies,
  ``QED effective action at finite temperature: Two loop dominance,''
  Phys.\ Rev.\ D {\bf 61}, 085021 (2000)
  [hep-ph/9909500].
%



\bibitem{Gies:2000ae} 
  H.~Gies,
  ``From effective actions to actual effects in QED,''
  AIP Conf.\ Proc.\  {\bf 564}, 68 (2001)
  [hep-ph/0010287].
\bibitem{Kim:2007ra} 
  S.~P.~Kim and H.~K.~Lee,
  ``Schwinger pair production at finite temperature in scalar QED,''
  Phys.\ Rev.\ D {\bf 76}, 125002 (2007)
  [arXiv:0706.2216 [hep-th]].
%
\bibitem{Kim:2008em} 
  S.~P.~Kim, H.~K.~Lee and Y.~Yoon,
  ``Schwinger Pair Production at Finite Temperature in QED,''
  Phys.\ Rev.\ D {\bf 79}, 045024 (2009)
  [arXiv:0811.0349 [hep-th]].
  %
\bibitem{Medina:2015qzc} 
  L.~Medina and M.~C.~Ogilvie,
  ``Schwinger Pair Production at Finite Temperature,''
  arXiv:1511.09459 [hep-th].

\bibitem{Selivanov:1985vt} 
  K.~B.~Selivanov and M.~B.~Voloshin,
``Destruction Of False Vacuum By Massive Particles,''
  JETP Lett.\  {\bf 42}, 422 (1985).

\bibitem{Selivanov:1986tu} 
  K.~G.~Selivanov,
  ``The Tunneling At Finite Temperature,''
  Phys.\ Lett.\ A {\bf 121}, 111 (1987).

\bibitem{Ivlev:1987zz} 
  B.~I.~Ivlev and V.~I.~Mel'nikov,
  ``Tunneling and activated motion of a string across a potential barrier,''
  Phys.\ Rev.\ B {\bf 36}, 6889 (1987).

\bibitem{Schutzhold:2008pz} 
  R.~Schutzhold, H.~Gies and G.~Dunne,
  ``Dynamically assisted Schwinger mechanism,''
  Phys.\ Rev.\ Lett.\  {\bf 101}, 130404 (2008)
  [arXiv:0807.0754 [hep-th]].
    %
\bibitem{Schneider:2014mla} 
  C.~Schneider and R.~SchŸtzhold,
  ``Dynamically assisted Sauter-Schwinger effect in inhomogeneous electric fields,''
  JHEP {\bf 1602}, 164 (2016)
  [arXiv:1407.3584 [hep-th]].


\bibitem{Kleban:2011cs} 
  M.~Kleban, K.~Krishnaiyengar and M.~Porrati,
  ``Flux Discharge Cascades in Various Dimensions,''
  JHEP {\bf 1111}, 096 (2011)
  [arXiv:1108.6102 [hep-th]].
    %
    %
    %
    %
    
    
\bibitem{Linde:1980tt} 
  A.~D.~Linde,
  ``Fate of the False Vacuum at Finite Temperature: Theory and Applications,''
  Phys.\ Lett.\  {\bf 100B}, 37 (1981);
  A.~D.~Linde,
  ``Decay of the False Vacuum at Finite Temperature,''
  Nucl.\ Phys.\ B {\bf 216}, 421 (1983)
  [Erratum-ibid.\ B {\bf 223}, 544 (1983)].
%



\bibitem{Coleman:1980aw} 
  S.~R.~Coleman and F.~De Luccia,
  ``Gravitational Effects on and of Vacuum Decay,''
  Phys.\ Rev.\ D {\bf 21}, 3305 (1980).
  
  %
\bibitem{Batra:2006rz} 
  P.~Batra and M.~Kleban,
  ``Transitions Between de Sitter Minima,''
  Phys.\ Rev.\ D {\bf 76}, 103510 (2007)
  [hep-th/0612083].
%

\bibitem{Brown:2007sd} 
  A.~R.~Brown and E.~J.~Weinberg,
  ``Thermal derivation of the Coleman-De Luccia tunneling prescription,''
  Phys.\ Rev.\ D {\bf 76}, 064003 (2007)
  [arXiv:0706.1573 [hep-th]].
%
%
%
%
%







\bibitem{Hackworth:2004xb} 
  J.~C.~Hackworth and E.~J.~Weinberg,
  ``Oscillating bounce solutions and vacuum tunneling in de Sitter spacetime,''
  Phys.\ Rev.\ D {\bf 71}, 044014 (2005)
  [hep-th/0410142].
  
\bibitem{Brown:2014rka} 
  A.~R.~Brown,
  ``Decay of hot Kaluza-Klein space,''
  Phys.\ Rev.\ D {\bf 90}, no. 10, 104017 (2014)
  [arXiv:1408.5903 [hep-th]].



\bibitem{Hawking:1981fz} 
  S.~W.~Hawking and I.~G.~Moss,
  ``Supercooled Phase Transitions in the Very Early Universe,''
  Phys.\ Lett.\ B {\bf 110}, 35 (1982).



\bibitem{Jensen:1983ac} 
  L.~G.~Jensen and P.~J.~Steinhardt,
  ``Bubble Nucleation and the {Coleman-Weinberg} Model,''
  Nucl.\ Phys.\ B {\bf 237}, 176 (1984).
  

  

    

    


\bibitem{Callan:1977pt} 
  C.~G.~Callan, Jr. and S.~R.~Coleman,
 ``The Fate of the False Vacuum. 2. First Quantum Corrections,''
  Phys.\ Rev.\ D {\bf 16}, 1762 (1977);
  G.~V.~Dunne, Q.~h.~Wang, H.~Gies and C.~Schubert,
  ``Worldline instantons. II. The Fluctuation prefactor,''
  Phys.\ Rev.\ D {\bf 73}, 065028 (2006)
  [hep-th/0602176].
%

    



  \bibitem{graphene}
  M.~I.~Katsnelson,  K.~S.~Novoselov, and A.~K.~Geim,
  ``Chiral tunnelling and the Klein paradox in graphene'', 
  Nature Physics 2, 620-625 (2006) 
  [cond-mat/0604323].

  
\bibitem{Allor:2007ei} 
  D.~Allor, T.~D.~Cohen and D.~A.~McGady,
  ``The Schwinger mechanism and graphene,''
  Phys.\ Rev.\ D {\bf 78}, 096009 (2008)
  [arXiv:0708.1471 [cond-mat.mes-hall]].
  %
  
  
\bibitem{Nieto:1985ws} 
  M.~M.~Nieto, V.~P.~Gutschick, F.~Cooper, D.~Strottman and C.~M.~Bender,
  ``Resonances In Quantum Mechanical Tunneling,''
  Phys.\ Lett.\  {\bf 163B}, 336 (1985).

  \bibitem{instability}
  J.~Plateau, ``Statique experimentale et theorique des liquides soumis aux seules forces moleculaires'', Gauthier-Villars, Ghent (1873);  L. Rayleigh, ``On the instability of jets'', Proceedings of the London Mathematical Society {\bf 1}, 4 (1878); 
  R.~Gregory and R.~Laflamme,
  ``Black strings and p-branes are unstable,''
  Phys.\ Rev.\ Lett.\  {\bf 70}, 2837 (1993)
  [hep-th/9301052].


  
\bibitem{Wang:1988ct} 
  R.~C.~Wang and C.~Y.~Wong,
  ``Finite Size Effect in the Schwinger Particle Production Mechanism,''
  Phys.\ Rev.\ D {\bf 38}, 348 (1988);
 %
 %
 %
 %
 %
 %
 %
 %
 %
  C.~Martin and D.~Vautherin,
  ``Finite Size Effects in Pair Production by an External Field,''
  Phys.\ Rev.\ D {\bf 38}, 3593 (1988).
  %
    

  %
  
  
    
\end{thebibliography}
\end{document}